\documentclass[12pt,english]{article}
\usepackage[utf8]{inputenc}
\usepackage{graphicx}
\usepackage{comment}
\usepackage{physics}
\usepackage{natbib}
\usepackage{amsmath}

\fontfamily{cmss}\selectfont
\usepackage{tgbonum}

\newcommand*{\crosssymbol}{%
    \text{%
      \raise 1ex\hbox{%
        \rlap{\vrule height.2pt depth.2pt width .75ex}%
        \hbox to .75ex{\hss\vrule height .5ex depth 1ex\hss}%
      }%
    }%
}
\makeatletter
\usepackage[utf8]{inputenc}
\usepackage{amssymb}

\usepackage[font=normalsize,font=doublespacing,labelfont=bf,justification=raggedright]{caption}

\usepackage{geometry}
\title{Does von Neumann Entropy Correspond to Thermodynamic Entropy?}
\author{Eugene Chua\thanks{\noindent I would like to thank Craig Callender, Eddy Keming Chen, Erik Curiel, Tim Maudlin, John Norton, Sai Ying Ng, Adrian K. Yee, and participants at the 2019 summer school \textit{The Nature of Entropy I} for their feedback, discussion and comments - they have contributed to this paper in one way or another. I would also like to thank two anonymous reviewers for pushing me to improve on various aspects of this paper. } \\ \textit{University of California San Diego} \\ \textit{Department of Philosophy} \\ \textit{eychua@ucsd.edu}}
\date{ }

\begin{document}
\maketitle

\begin{abstract}
\noindent Conventional wisdom holds that the von Neumann entropy corresponds to thermodynamic entropy, but Hemmo and Shenker (2006) have recently argued against this view by attacking von Neumann's (1955) argument. I argue that Hemmo and Shenker’s arguments fail due to several misunderstandings: about statistical-mechanical and thermodynamic domains of applicability, about the nature of mixed states, and about the role of approximations in physics. As a result, their arguments fail in all cases: in the single-particle case, the finite particles case, and the infinite particles case. 
\end{abstract}

\section{Introduction}

According to conventional wisdom in physics, von Neumann entropy corresponds to phenomenological thermodynamic entropy. The origin of this claim is von Neumann's (1955) argument that his proposed entropy corresponds to the thermodynamic entropy, which appears to be the only explicit argument for the equivalence of the two entropies. However, Hemmo and Shenker (\textbf{H$\&$S}) (2006) -- and earlier, Shenker (1999) -- have argued that this correspondence fails, contrary to von Neumann. If so, this leaves conventional wisdom without explicit justification.

Correspondence can be understood, at the very least, as a numerical consistency check: in this context, this means that the von Neumann entropy has to be included in calculating thermodynamic entropy to ensure consistent accounting in contexts where both thermodynamic and von Neumann entropy are physically relevant. Successful correspondence provides strong evidence of equivalence. While it does not guarantee equivalence, it seems to be at least a necessary condition for equivalence. If thermodynamic entropy and von Neumann entropy correspond, then we have reason to think that von Neumann entropy is rightfully thermodynamic in nature, since proper accounting of thermodynamic entropy would demand von Neumann entropy. By contrast, a failure of correspondence seems to entail that the von Neumann entropy is \textit{not} thermodynamical in nature, since it is irrelevant to thermodynamic calculations in contexts where both entropies are physically significant (e.g. when a system has both quantum degrees of freedom \textit{and} is sufficiently large to warrant thermodynamical considerations). 

Although Henderson (2003), in my view, has successfully criticized Shenker's earlier argument, little has been done in the philosophical literature to evaluate \textbf{H$\&$S}'s more recent arguments.\footnote{It is only slightly better in the physics literature: Deville and Deville (2013) appears to be the only paper to critique \textbf{H$\&$S}. On the philosophical side, one (very recent) exception is Prunkl (ms), though she restricts discussion to the single-particle case and appears to conflate information entropy with thermodynamic entropy. See $\S$4.1/$\S$4.2 for why this is not obviously right.} This lacuna is striking because, as I mentioned, von Neumann’s argument appears to be the only explicit argument for correspondence for the two entropies.

My goal in this paper is to fill this lacuna by providing a novel set of criticisms to \textbf{H$\&$S}. Here's the plan: I introduce key terms ($\S$2) and then present von Neumann's thought-experiment which aims to establish the correspondence between thermodynamic entropy and von Neumann entropy; along the way, a novel counterpart to the usual argument for correspondence is discussed ($\S$3). I then present and criticize \textbf{H$\&$S}'s arguments for the single-particle case in the context of thermodynamics ($\S$4.1) and in the context of statistical mechanics ($\S$4.2), the N-particles case ($\S$4.3), and the infinite-particles case ($\S$4.4). I conclude that their argument fails in all cases -- in turn, we have good reasons to reject their claim that the von Neumann entropy fails to correspond to thermodynamic entropy, and hence the claim that von Neumann entropy is not thermodynamic in nature.

\section{Key Terms}

Let me first define the notions of thermodynamic entropy and von Neumann entropy. Following \textbf{H$\&$S}, I define the \textit{change in} \textit{thermodynamic entropy} \textit{S}$_{\text{TD}}$ between two thermodynamic states in an \textit{isothermal quasi-static} process,\footnote{There is no change in temperature in an isothermal quasi-static process, which is why $T$ is taken to be constant. As a matter of historical note, von Neumann uses an isothermal set-up in his argument, with a box containing a quantum ideal gas coupled to a (much larger) heat sink ensuring constant temperature over time (von Neumann 1955, 361/371).} as:
\begin{equation}
\Delta S_{TD} = \frac{1}{T}\int P\text{ }dV 
\end{equation}

\noindent We will restrict our discussion to ideal gases in equilibrium (i.e. systems where pressure $P$, volume $V$, and temperature $T$ remain constant). 

Next, the \textit{von Neumann entropy} \textit{S}$_{\text{VN}}$, for any pure or mixed quantum system, is defined as:
\begin{equation}
\textit{S}_{\text{VN}} = -kTr(\rho\text{ }log\text{ }\rho)
\end{equation}

\noindent where $k$ is the Boltzmann constant and Tr(.) is the trace function. Generally, the density matrix $\rho$ is such that: 
\begin{equation}
\rho = \sum\limits_{n=1}^{i} p_i \ket{{\psi}_i}\bra{{\psi}_i}
\end{equation}

\noindent where ${\psi}_1, {\psi}_2, ...\text{ } {\psi}_n$ correspond to the number of pure states in a statistical mixture represented by $\rho$, with $p_1, p_2, ... \text{ } p_n$ being their associated classical probabilities (which must sum to unity). In the case where there is only one pure state possible for a system (e.g. when we are absolutely certain about its quantum state), then $n$ = 1, with probability 1, so the appropriate density matrix is $\rho = \ket{{\psi}}\bra{{\psi}}$. For such a system in a pure state (i.e. represented by a single state vector in Hilbert space), $S_{\text{VN}} = 0$. For mixed states (i.e. states which cannot be represented by a single state vector in Hilbert space, hence \textit{mixture} of pure states or a \textit{mixed state}), Tr($\rho$ log $\rho$) $<$ 1 and $S_{\text{VN}} > 0$ in general. A mixed state is often said to represent our \textit{ignorance} about a system -- this will suffice as a first approximation (more on how to interpret this ignorance in $\S$4.2).

\textit{Prima facie}, \textit{S}$_{\text{VN}}$ and \textit{S}$_{\text{TD}}$ appear to share nothing in common, apart from the word `entropy'. However, von Neumann claims that there are important correlations between the two, which suggests a correspondence between \textit{S}$_{\text{TD}}$ and \textit{S}$_{\text{VN}}$.

\section{Von Neumann's Thought-Experiment}

For the sake of parity, I adopt \textbf{H$\&$S}'s presentation of von Neumann's thought-experiment,\footnote{It is not clear to me that von Neumann's original 1932/1955 argument is exactly the same as the argument \textbf{H$\&$S} reproduces. However, for the sake of argument, I will refer to \textbf{H$\&$S}'s version as von Neumann's argument in this paper.}  which aims to show that changes in thermodynamic entropy can only be made consistent with the laws of thermodynamics if we considered the von Neumann entropy as contributing to the calculation of the thermodynamic entropy. \textit{Fig. 1.} depicts the stages of the thought-experiment.

We begin, in stage one, with a box with a partition in the middle. On one side of the partition there is a gas at volume $V$, constant temperature $T$, and constant pressure $P$. Each gas particle starts off having the pure state spin-up along the x-direction $\ket{\psi_x^{\uparrow}}$, which is equivalent to a superposition of spin-up and spin-down pure states along the z-direction, labelled $\ket{\psi_z^\uparrow}$ and $\ket{\psi_z^\downarrow}$ respectively. According to standard quantum mechanics, the state of each particle is thus $\frac{1}{\sqrt2}(\ket{\psi_z^\uparrow} + \ket{\psi_z^\downarrow})$. 

In this context, particles with quantum behavior may be taken to be \textit{ideal gases}, i.e. sets of particles each of which do not interact with other particles and take up infinitesimal space. Following von Neumann's assumptions (von Neumann 1955, 361),\footnote{These assumptions are borrowed from Einstein (1914). For more, see Peres (Peres 2002, 271).} each gas particle is understood as a quantum particle with a spin degree of freedom contained inside a large impenetrable box, and each gas particle is put inside an even larger container isolated from the environment (i.e. the box we began with). This ensures that each spin degree of freedom is incapable of interacting with other particles. These boxes' sizes also ensure that the positions of these boxes (and hence of the particles) can be approximately classical. Since the container is much larger than each gas particle, this ensures that the gas particles take up negligible space relative to the massive container. Accepting these assumptions, we may then take these quantum particles to behave like an ideal gas.\footnote{I shall follow everyone in this debate in assuming that the above set-up is physically possible.} Following \textbf{H$\&$S}, we further assume that the position degrees of freedom of the gas particles have no interaction with the spin degrees of freedom at this point, and ``due to the large mass of the boxes, the position degrees of freedom of the gas may be taken to be classical and represented by a quantum mechanical mixture''. (Hemmo and Shenker 2006, 155)

Moving on, stage two involves a spin measurement along the z-axis on all the particles in the container, with a result being an equally weighted statistical mixture of particles with either $\ket{\psi_z^\uparrow}$ or $\ket{\psi_z^\downarrow}$ states. As a result, the spin state of each particle is then represented instead by a density matrix $\rho_{spin}$, such that: 

\begin{equation}
\rho_{spin} = \frac{1}{2}(\ket{\psi_z^\uparrow}\bra{\psi_z^\uparrow} + \ket{\psi_z^\downarrow}\bra{\psi_z^\downarrow})
\end{equation}

\noindent More precisely, there should be terms for the measurement device too, when truly considering the entire system. $\rho_{spin}$ describes only the subsystem (i.e. the quantum ideal gas) \textit{sans} measurement device, i.e. a state with the measurement device traced out -- this is in line with von Neumann's focus on the entropy changes due to changes in the subsystem (von Neumann 1955, 358--379). I follow Henderson (2003) and \textbf{H$\&$S} in talking about the system's state as though I have already traced the measurement device out whenever measurement is involved. 

Stages three and four are where the particles are (reversibly) separated according to their spin states by a semi-permeable wall into two sides of the box, each with volume $V$.\footnote{This semi-permeable wall can be assumed to be a black box which reversibly separates particles to different sides based on their different orthogonal/disjoint states; see (von Neumann 1955, 367--370) for discussion. I follow everyone in the debate in accepting this assumption.} As a result of this separation, we in effect double the mixture's volume. The gas expands to fill up volume $V$ on each side.
 
Stage five involves an isothermal and quasi-static compression of the mixture so that we return to a total volume $V$ (effectively halving the volume on each side of the box), while pressure on both sides becomes equal. Importantly, due to this compression, \textit{S}$_{\text{TD}}$ decreases due to the decrease in volume. 

Stage six brings all the particles into the pure spin state $\ket{\psi_x^{\uparrow}}$ quasi-statically and without work done, while stage seven removes the semi-permeable wall, such that the system returns to its original state. 

\begin{center}
    \includegraphics[scale=0.20]{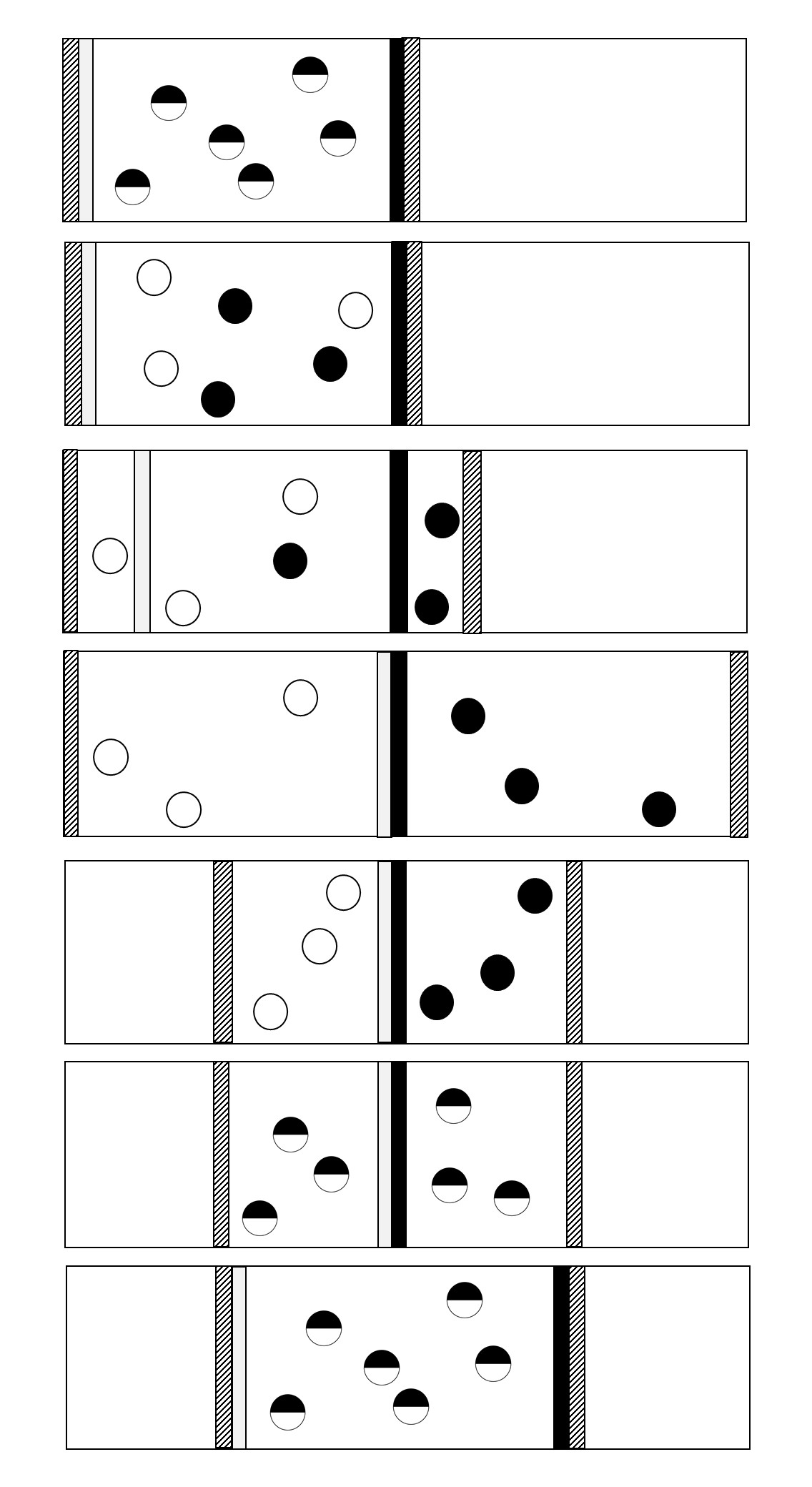}
    
    \footnotesize{\textit{Fig. 1: from top to bottom, stage one to stage seven, as described by \textbf{H$\&$S}.}}
\end{center}

Now consider how $S_{\text{VN}}$ and $S_{\text{TD}}$ change across the various stages. Stage seven ends with the body of gas having the same thermodynamic state (same $V$, same $P$, and constant $T$) as stage one. Furthermore, all the thermodynamic transformations performed were reversible, and removing the wall alone does no additional work. Thus, the system at stage one must have the same thermodynamic entropy as stage seven, i.e. $\Delta S_{\text{TD}}$ = 0, since \textit{S}$_{\text{TD}}$ depends only on the initial and final state of the system. $\Delta S_{\text{VN}} = 0$ from stage one to seven too, since the system is in the same state in both the first and seventh stages.

Since stage six does not involve thermodynamic transformations, there is no change in \textit{S}$_{\text{TD}}$. Likewise, the transformation of $\rho_{spin}$ to $\ket{\psi_x^{\uparrow}}$ here does not change \textit{S}$_{\text{VN}}$ as the transformation can be performed unitarily. This is possible as a result of our separation of the gases to different sides of the box according to their spin-eigenstates - given this, we can perform unitary operations on each side of the box (or perform the more general measurement procedure recommended by (von Neumann 1955, 365-367)), to transform them into the same state as stage one. Both unitary transformations and von Neumann's procedure do not increase \textit{S}$_{\text{VN}}$, and so there is no change in \textit{S}$_{\text{VN}}$ at stage six as a result.

There are no changes in \textit{S}$_{\text{TD}}$ or \textit{S}$_{\text{VN}}$ in stages three and four. While there is an increase in the gas's volume, as noted above, from $V$ to $2V$, and hence an accompanying increase in \textit{S}$_{\text{TD}}$ by $n.R.log\textit{ }2$,\footnote{Here, $n$ refers to the number of moles of gas in the system, and $R$ is the gas constant.} there is also a compensating change in the thermodynamic entropy of mixing\footnote{Henderson (2003) explains the mixing entropy, describing the mixing of different gases, crisply: ``After separation, each separated gas occupies the original volume $V$ alone. To return to the mixture, each gas is compressed to a volume $c_{i}V$ (where $c$ is the concentration of the $i^{th}$ gas). The compression requires work $W = -n.k.T\sum_{i}\textit{ }c_{i}\textit{ }log\textit{ }c_{i}$ to be invested, and the entropy of the gas is reduced by $\Delta S = -n.k.\sum_{i}\textit{ }c_{i}\textit{ }log\textit{ }c_{i}$. An increase in entropy of the same amount must then be associated with the mixing step of removing the partitions. This is the `mixing entropy'.'' (Henderson 2003, 292) Separation simply results in a decrease in entropy of the same amount.}$^{,}$\footnote{Tim Maudlin raised the following objection to the applicability of the entropy of mixing in this context when a version of this paper was presented at a summer school. Mixing should have a thermodynamic effect only when differences between the gases are already assumed to be thermodynamically relevant: for example, mixing differently colored gases should not have a thermodynamic effect unless the difference in color is thermodynamically relevant. It is, however, not clear whether the difference in spin is a thermodynamically relevant one, and might amount to begging the question. This is a good point, but one that I am setting aside for now, since everyone in the debate accepts the assumption that separating the gases here decreases the entropy of mixing. As we shall see later, a more fundamental issue arises with using the entropy of mixing in the `single particle' case.} by $-n.R.log\textit{ }2$ which exactly compensates this increase in \textit{S}$_{\text{TD}}$.\footnote{see (\textbf{H$\&$S} 2006, 157, fn. 4).} Since the particles are in orthogonal spin states at this stage, there are no quantum effects (e.g. `collapse' effects) from simply filtering the gases with the semi-permeable walls, and hence \textit{S}$_{\text{VN}}$ does not change either.\footnote{This is argued for in (von Neumann 1955, 370--376).}

However, importantly, there is a decrease in \textit{S}$_{\text{TD}}$ in stage five, of $-n.R.log\textit{ }2$ due to the isothermal compression and decrease in volume. Yet, nowhere else is there any further change in \textit{S}$_{\text{TD}}$. We have to account for why the overall change in \textit{S}$_{\text{TD}}$ from the first to the seventh stages is 0. 

As von Neumann argues, only one possibility remains. While \textit{S}$_{\text{TD}}$ remains constant in stage two, notice that there was an increase in \textit{S}$_{\text{VN}}$, of $-N.k.-log\textit{ }2 = \frac{N.R}{N_{A}}.log\textit{ }2 = n.R.log\textit{ }2$,\footnote{$N$ is the total number of particles: since each particle is assumed to be non-interacting and independent from others under the ideal gas assumption, their entropies are additive. $N_A$ is Avogadro's number.}  as a result of the spin measurement. This is equivalent to the change of \textit{S}$_{\text{TD}}$ in stage five. The state of each particle changes from a pure state $\frac{1}{\sqrt2}(\ket{\psi_z^\uparrow} + \ket{\psi_z^\downarrow})$ to a mixed state represented by $\rho_{spin}$, and hence \textit{S}$_{\text{VN}}$ for the gas increases on the whole. In order to ensure that entropic changes are consistent, von Neumann thinks that we should accept \textit{S}$_{\text{VN}}$'s contribution to \textit{S}$_{\text{TD}}$ in this context, where both quantum effects and thermodynamical considerations are at play. Without accepting \textit{S}$_{\text{VN}}$ in our entropic accounting, we end up with a violation of thermodynamics since we have a reversible thermodynamic cycle with non-zero change in \textit{S}$_{\text{TD}}$, \textit{contra} the Second Law. In other words, we should accept that \textit{S}$_{\text{VN}}$ corresponds to \textit{S}$_{\text{TD}}$.

Furthermore, the correspondence of \textit{S}$_{\text{VN}}$ and \textit{S}$_{\text{TD}}$ in this context can be defended from another perspective, apart from considerations about consistency from the thermodynamic perspective: consistent accounting from the perspective of \textit{quantum mechanics} also demands correspondence. This is simply a change in perspective with regards to the thought-experiment, but, to my knowledge, this argument has not been explicitly made in the literature, thus underselling the case for correspondence in von Neumann's thought experiment. 

Instead of arguing for correspondence by considering thermodynamic consistency, i.e.  ensuring that $\Delta S_{\text{TD}}$ = 0 throughout the cycle, we can also consider consistency from the quantum mechanical perspective. We started and ended with the same spin state, and so it should be the case that $\Delta S_{\text{VN}} = 0$ throughout the cycle. Yet, there is an inconsistency: if we only consider the increase of \textit{S}$_{\text{VN}}$ in stage two as a result of measurement, we should end in stage seven with an \textit{increase} in \textit{S}$_{\text{VN}}$, \textit{not} $\Delta S_{\text{VN}} = 0$. As described, there is nowhere else in the thought-experiment where \textit{S}$_{\text{VN}}$ changes. However, there \textit{is} a decrease in \textit{S}$_{\text{TD}}$ in stage five due to the \textit{thermodynamic process} of isothermal compression, exactly balancing out the increase in \textit{S}$_{\text{VN}}$. Hence, we can ensure consistency, i.e. that $\Delta S_{\text{VN}} = 0$, only by taking \textit{S}$_{\text{VN}}$ to correspond to \textit{S}$_{\text{TD}}$. In other words, just as the thermodynamic accounting of \textit{S}$_{\text{TD}}$ is consistent only if we consider \textit{S}$_{\text{VN}}$, the quantum entropic accounting of \textit{S}$_{\text{VN}}$ is \textit{also} consistent only if we consider \textit{S}$_{\text{TD}}$. Consistency from a quantum mechanical perspective also demands correspondence between \textit{S}$_{\text{VN}}$ and \textit{S}$_{\text{TD}}$. 

Though the debate has largely focused only on how the thought-experiment demonstrates one direction of correspondence, of \textit{S}$_{\text{VN}}$ \textit{to} \textit{S}$_{\text{TD}}$ as a result of thermodynamical considerations, the correspondence demonstrated by this thought-experiment in fact goes \textit{both ways}. Of course, since von Neumann was focused on demonstrating the \textit{thermodynamic} nature of $S_{\text{VN}}$ (specifically the irreversibility of measurement), rather than the \textit{quantum} nature of \textit{S}$_{\text{TD}}$, it was natural that he chose to approach it the way he did.

\section{Hemmo and Shenker's Arguments}

\textbf{H$\&$S} disagree with von Neumann's argument, and criticize it by considering three cases: the single-particle case, the finite but large $N$ particles case, and the infinite particles case.

\subsection{Single Particle Case - Thermodynamics}

\textbf{H$\&$S} first consider von Neumann's argument in the single particle case (see Fig. 2). They claim that the argument does not go through here, since \textit{S}$_{\text{TD}}$ actually remains constant, contrary to our thought-experiment's description. In other words, using thermodynamical considerations, they find that \textit{S}$_{\text{VN}}$ should not be included in our accounting for \textit{S}$_{\text{TD}}$.

Here's their argument. Consider the stages where there are entropic changes. In stage two when the spin measurement was performed, \textit{S}$_{\text{VN}}$ increases as before, since it tracks the change of the particle's spin state from pure to mixed. 

Contrariwise, \textit{S}$_{\text{TD}}$ does not change in stage five (isothermal quasi-static compression) \textit{nor anywhere else} (this will be important later). After stage two, the single particle is in either the $\ket{\psi_z^\uparrow}$ state or the $\ket{\psi_z^\downarrow}$ state. After stages three and four, with the expansion and separation via semi-permeable wall, there is a particle only in \textit{one} side of the box, and not the other. We make an \textit{S}$_{\text{TD}}$-conserving location measurement\footnote{Prunkl (ms) claims that the location measurement leads to a violation of the Second Law. If true, this makes \textbf{H$\&$S}'s argument even more problematic. Here, for the sake of argument, I assume that the location measurement is unproblematic.} to figure out which side of the box is empty and which side the particle is at, so as to compress the box against the empty side. The compression is then performed as per before. However, this compression does \textit{not} decrease \textit{S}$_{\text{TD}}$:\footnote{As an anonymous reviewer rightfully notes, the location measurement is important for ensuring $\Delta$\textit{S}$_{\text{TD}} = 0$ here. Without the location measurement, we might end up compressing in the wrong direction against the side with the gas, rather than the empty vacuum - this will have thermodynamic effects since we are doing work on the gas. However, the \textbf{H$\&$S} set-up emphasizes the location measurement, and I will play along for the sake of argument.} to restore the volume of the `gas' to $V$ no work needs to be done, since we are compressing against vacuum. Since there is a change in \textit{S}$_{\text{VN}}$ in this cycle, but no change in \textit{S}$_{\text{TD}}$, the apparent answer, in order to do our entropic accounting, is to \textit{ignore}, not incorporate, \textit{S}$_{\text{VN}}$ into \textit{S}$_{\text{TD}}$. Hence \textit{S}$_{\text{VN}}$ does not correspond to \textit{S}$_{\text{TD}}$. 

\begin{center}
    \includegraphics[scale = 0.2]{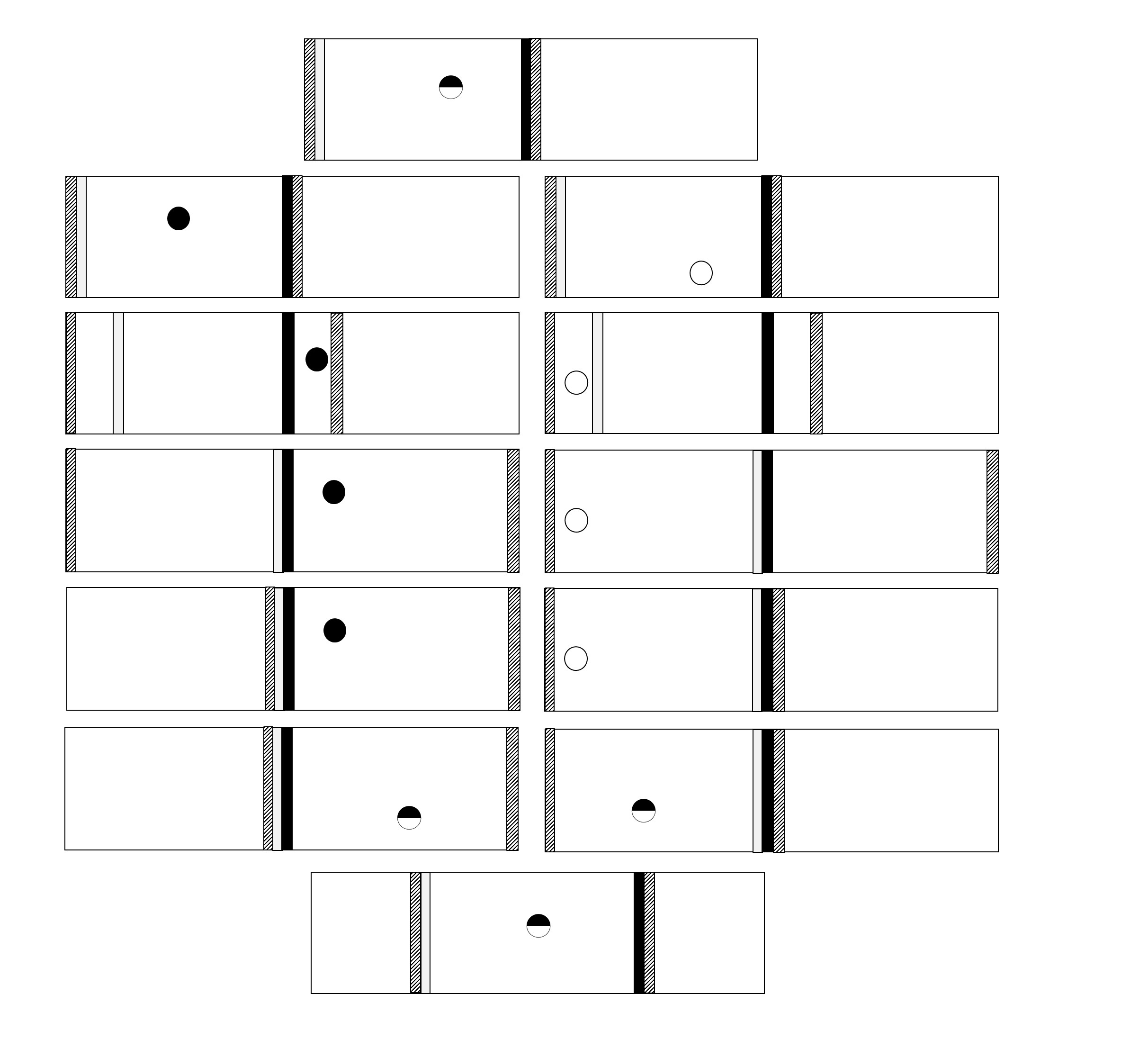}
    
    \footnotesize{\textit{Fig. 2: from top to bottom, stage one to stage seven for the single particle case as described by \textbf{H$\&$S}.}}
\end{center}

\noindent Their analysis is problematic. Though their ultimate point in this analysis -- that $S_{\text{TD}}$ fails to corresponds to $S_{\text{VN}}$ -- still holds, it does not hold in the way they claim. In fact, the way it fails suggests to us that we should \textit{disregard} the single particle case. 

For the single particle case, they claim that ``... [\textit{S}$_{\text{TD}}$] is null throughout the experiment." (\textbf{H$\&$S} 2006, 162) This then allows them to claim that thermodynamic accounting for \textit{S}$_{\text{TD}}$ is consistent \textit{only if} we did not consider \textit{S}$_{\text{VN}}$. This then supports their claim that \textit{S}$_{\text{VN}}$ does not correspond to \textit{S}$_{\text{TD}}$ since adding \textit{S}$_{\text{VN}}$ into the thermodynamic accounting actually renders the otherwise consistent calculations inconsistent.

They are right to say that the stage five compression (after location measurement) has no thermodynamic effect because we are compressing against vacuum: no work needs to be done, and so $\Delta$\textit{S}$_{\text{TD}} = 0$ for stage five. \textit{However}, I claim that $\Delta$\textit{S}$_{\text{TD}} \neq 0$ for the single particle case \textit{overall}, because $\Delta$\textit{S}$_{\text{TD}} \neq 0$ in stages three and four in this context. 

As far as I can tell, \textbf{H$\&$S} did not analyze stage three and four, i.e. the isothermal expansion and separation, in terms of the single particle case at all. Rather, they seem to have assumed that $\Delta$\textit{S}$_{\text{TD}} = 0$ in these stages as with the original case of the macroscopic gas.\footnote{Prunkl (ms) appears to do the same.} However, this assumes that there is both a change in entropy of $n.R.log\textit{ }2$ due to isothermal expansion \textit{and} a change in the entropy of mixing of $-n.R.log\textit{ }2$ due to separation, as they say so themselves for the original case: ``The increase of thermodynamic entropy due to the volume increase $\Delta$\textit{S} = $\frac{1}{T}\int P\text{ }dV$ is exactly compensated by the decrease of thermodynamic mixing entropy $\Delta$\textit{S} = $\sum w_k\textit{ }ln\textit{ }w_k$ (where $w_k$ is \textit{the relative frequency of molecules of type k}) due to the separation." (\textbf{H$\&$S} 2006, 157, fn. 4, emphasis mine)

In the single particle case, it makes sense that isothermal expansion should still increase \textit{S}$_{\text{TD}}$, since the single particle `gas' is expanding against a piston and doing work. However, it \textit{does not make physical sense} to speak of the entropy of mixing here at all, since \textit{there is no separation of gases} in the single particle case. The entropy of mixing is explicitly defined for systems where \textit{different} gases are separated from/mixed with one another via semi-permeable walls, but a single particle cannot be separated from/mixed with itself. The quote above makes this conceptual point explicit: by \textbf{H$\&$S}'s own lights, the relative frequency of a single particle is simply unity (and null for particles of other types), so the entropy of mixing is $1\textit{ }ln\textit{ }1 = 0$. \textit{There is no thermodynamic entropy of mixing in the single particle case}.

Discounting the entropy of mixing, however, we find that $\Delta$\textit{S}$_{\text{TD}} = n.R.log\textit{ }2 \neq 0$ for stages three and four, and hence for the entire process, contrary to \textbf{H$\&$S}'s claim. Interestingly, correspondence \textit{does} fail to obtain between \textit{S}$_{\text{TD}}$ and \textit{S}$_{\text{VN}}$, since $\Delta$\textit{S}$_{\text{TD}} +$ $\Delta$\textit{S}$_{\text{VN}} = 2n.R.log\textit{ }2 \neq 0$, despite the process being reversible \textit{ex hypothesi}: incorporating \textit{S}$_{\text{VN}}$ into thermodynamic accounting violates the Second Law.

However, on this new analysis, we gain some clarity as to why the single particle case is problematic. While it is true that incorporating \textit{S}$_{\text{VN}}$ into the thermodynamic accounting violates the Second Law, \textit{S}$_{\text{TD}}$ accounting \textit{by itself} also violates the Second Law (contrary to \textbf{H$\&$S}). Even without considering \textit{S}$_{\text{VN}}$, $\Delta$\textit{S}$_{\text{TD}} \neq 0$ despite the process being reversible. Thermodynamic accounting is inconsistent here \textit{no matter what we do}, which suggests that the reversible process they described for the single particle case is thermodynamically unsound: if so, any argument \textbf{H$\&$S} make in this context may be disregarded.

The upshot: I agree with \textbf{H$\&$S} that correspondence fails for the single particle case, but not \textit{why} it fails. It is \textit{not} because the process they described is already thermodynamically consistent without taking \textit{S}$_{\text{VN}}$ into account. Rather, it is because the process is already thermodynamically \textit{inconsistent} anyway.

In recent work, John Norton argued that thermodynamically reversible processes for single-particle systems are impossible in principle, which might explain \textit{why} the process described by \textbf{H$\&$S} is thermodynamically unsound: it was not justified to assume the process was reversible for a single particle system. For Norton, a reversible process is ``loosely speaking, one whose driving forces are so delicately balanced around equilibrium that only a very slight disturbance to them can lead the process to reverse direction. Because such processes are arbitrarily close to a perfect balance of driving forces, they proceed arbitrarily slowly while their states remain arbitrarily close to equilibrium states." (Norton 2017, 135) Norton notes that these thermodynamic equilibrium states are balanced not because there are no fluctuations, but because these fluctuations are negligible for macroscopic systems. However, fluctuations relative to single-particle systems are large, and generally prevent these systems from being in equilibrium states at any point of the process, rendering reversible processes impossible in the single particle case. (Norton 2017, 135) If reversible processes are impossible for single particle systems in general, then it should come as no surprise that the particular single particle reversible process used by \textbf{H$\&$S} is likewise thermodynamically unsound, as my analysis above suggests. If so, \textbf{H$\&$S}'s claim that correspondence fails in this process is simply besides the point, since this process is not thermodynamic at all. 

Since any reversible process cannot be realized for single particle systems in general, the issue seems not to be with any particular process \textit{per se}, but with the single particle case \textit{simpliciter}. To my knowledge, no one prior to \textbf{H$\&$S} discussed von Neumann's experiment in terms of a single particle; von Neumann (1955), Peres (1990, 2002), Shenker (1999) and Henderson (2003) all explicitly or implicitly assume a large (or infinite) number of particles. This is for good reason. As \textbf{H$\&$S} acknowledge, and as we have seen: ``The case of a single particle is known to be problematic as far as arguments in thermodynamics are concerned''. (\textbf{H$\&$S} 2006, 158) Matter in phenomenological thermodynamics is assumed to be continuous.\footnote{See Compagner (1989) for a discussion of the so-called `continuum limit' as a counterpart to the thermodynamic limit in phenomenological thermodynamics.} A `gas' composed of one particle can be many things, but it is surely not continuous in any commonly accepted sense. In other words, it is just not clear whether the domain of thermodynamics should apply to the single-particle case at all. 

As Myrvold (2011) notes, Maxwell also made a similar claim with regards to phenomenological thermodynamics \textit{in general}; it does not and should not hold in the single particle case. On his view, the laws of phenomenological thermodynamics, notably the Second Law, must be continually violated on small scales:

\begin{quote}

If we restrict our attention to any one molecule of the system, we shall find its motion changing at every encounter in a most irregular manner. 

If we go on to consider a finite number of molecules, even if the system to which they belong contains an infinite number, the average properties of this group, though subject to smaller variations than those of a single molecule, are still every now and then deviating very considerably from the theoretical mean of the whole system, because the molecules which form the group do not submit their procedure as individuals to the laws which prescribe the behaviour of the average or mean molecule.

Hence the second law of thermodynamics is continually being violated, and that to a considerable extent, in any sufficiently small group of molecules belonging to a real body. As the number of molecules in the group is increased, the deviations from the mean of the whole become smaller and less frequent [...] (Maxwell 1878, 280)
\end{quote}

\noindent The Second Law, and hence phenomenological thermodynamics, should not be expected to hold true universally in small scale cases, and especially \textit{not} in the single-particle case. Von Neumann and everyone else in the debate should have recognized this point. Why, then, should it matter that the thought-experiment succeeds or fails in this case? Phenomenological thermodynamics does not apply to single-particle cases. There is thus no profit in trying to establish correspondence between \textit{S}$_{\text{VN}}$ and \textit{S}$_{\text{TD}}$ in this case. Indeed, if we took seriously Maxwell's claim that the Second Law fails at small scales, a failure of thermodynamic entropic accounting might even be \textit{expected}; it does not rule out the possible thermodynamic nature of \textit{S}$_{\text{VN}}$ even though the sum of \textit{S}$_{\text{VN}}$ and \textit{S}$_{\text{TD}}$ might be inconsistent with the Second Law. In short, it is not clear \textit{why} the single-particle case is relevant to the discussion at hand.

\textbf{H$\&$S}'s reasoning is untenable, because they fail to respect the context of phenomenological thermodynamics by bringing it into a context where it is not expected to hold. Instead, it seems more appropriate that the single-particle case is precisely beyond the purview of classical thermodynamics, requiring an analogue that only corresponds to classical thermodynamics at the appropriate scales and limits. We may then take \textit{S}$_{\text{VN}}$ to be the analogue of \textit{S}$_{\text{TD}}$ in this case, only approximating \textit{S}$_{\text{TD}}$ as the system in question approaches the context suitable for traditional thermodynamic analysis. If so, we may see von Neumann as merely demonstrating that \textit{S}$_{\text{VN}}$ corresponds, not at \textit{all} domains but \textit{in the domain where thermodynamics is taken to hold}, to \textit{S}$_{\text{TD}}$.

\subsection{Single Particle Case Redux - Statistical Mechanics and Information}

Given the foregoing discussion, \textbf{H$\&$S} might insist that \textit{S}$_{\text{VN}}$ fails to correspond to \textit{S}$_{\text{TD}}$ \textit{even when} take into account a more relevant domain for single particles -- statistical mechanics.

After directly arguing that \textit{S}$_{\text{VN}}$ does not correspond to \textit{S}$_{\text{TD}}$ (\textbf{H$\&$S} 2006, 162--165), they further argue that \textit{S}$_{\text{VN}}$ does not correspond to information entropy (more on this below) in the single-particle case. \textit{Prima facie}, this should seem \textit{irrelevant} to von Neumann's argument, which was to establish the correspondence of the \textit{thermodynamic} \textit{S}$_{\text{TD}}$ and \textit{quantum} \textit{S}$_{\text{VN}}$: why should \textit{information} entropy's failure to correspond with \textit{S}$_{\text{VN}}$ be a worry at all? 

Here's one plausible worry, on a charitable reading. \textit{If} information entropy corresponds to \textit{S}$_{\text{TD}}$, \textit{and} \textbf{H$\&$S} shows that \textit{S}$_{\text{VN}}$ fails to correspond to information entropy, then we might conclude, indirectly, that \textit{S}$_{\text{VN}}$ does not correspond to \textit{S}$_{\text{TD}}$ after all.\footnote{Caveat: I am not committed to the information entropy's relationship to thermodynamics. One may, like Earman and Norton (1998, 1999), be skeptical that information entropy is related to \textit{S}$_{\text{TD}}$ at all, in which case \textbf{H$\&$S}'s argument here is simply irrelevant.} This argument assumes that information entropy \textit{does} correspond to \textit{S}$_{\text{TD}}$, an assumption \textbf{H$\&$S} seem to hold as well: this is in line with the so-called `subjectivist' view of statistical mechanics.\footnote{Notably, see Jaynes (1957).}  Furthermore, my above argument against the misapplication of phenomenological thermodynamics does not seem to apply here, since this argument is being made in the context of statistical mechanics and its particle picture, with no commitment to phenomenological thermodynamics.

However, \textbf{H$\&$S} do not do much to motivate the linkage between information entropy and \textit{S}$_{\text{TD}}$; indeed, in their words, ``a linkage between the Shannon information and thermodynamic entropy has not been established'' (\textbf{H$\&$S} 2006, 164). Without this link, the failure of correspondence between the information entropy and \textit{S}$_{\text{VN}}$ appears, at best, irrelevant to the correspondence between \textit{S}$_{\text{TD}}$ and \textit{S}$_{\text{VN}}$. Nevertheless, I will take a charitable view here and assume that there \textit{is} a correspondence between information entropy and \textit{S}$_{\text{TD}}$, for the sake of assessing their argument. Here's a plausible (if arguable) sketch: if one were a subjectivist like Jaynes (1957), one might take the Gibbs entropy in statistical mechanics to be a special case of the information entropy. After all, both have the form: 
\begin{equation}
-\sum_{i} p_i\text{ }ln\text{ }p_i
\end{equation}

\noindent with \textit{i} being the number of possible states with associated probabilities of occurring $p_i$, with the Gibbs entropy being multiplied by an additional Boltzmann's constant $k$.\footnote{Using the so-called Planck units, where $k = 1$, Gibbs entropy and information entropy are then formally equivalent.} We know that statistical mechanics corresponds to phenomenological thermodynamics at the thermodynamic limit so we can think of the Gibbs entropy, and hence information entropy, as corresponding to \textit{S}$_{\text{TD}}$. I take this to be in line with what \textbf{H$\&$S} have in mind: ``to the extent that the Shannon information underwrites the thermodynamic entropy, it does so via statistical mechanics'' (2006, 165). \textit{Assuming} that the above picture is plausible, a failure of correspondence between \textit{S}$_{\text{VN}}$ and the information entropy provides evidence against the correspondence between \textit{S}$_{\text{VN}}$ and \textit{S}$_{\text{TD}}$. 

Their argument comes into two parts. Ignoring \textit{S}$_{\text{TD}}$ for the time being (which does not change throughout the cycle for the single-particle case -- see $\S$4.1), they claim that we can consider the stage five location measurement to be a decrease in \textit{information entropy} of $ln$ $2$, as a result of learning information about which one of two parts of the box the particle is in. On first glance, this seems to resolve the arithmetic inconsistency in entropic accounting: $ln$ 2 is exactly the increase in \textit{S}$_{\text{VN}}$ as a result of the spin state changing from a pure state $\ket{\psi_x^{\uparrow}}$ to the mixed state $\rho_{spin}$ in stage two. In other words, for both the information and von Neumann entropy's accounting to be correct (i.e. net change of zero across the cycle), we must consider \textit{S}$_{\text{VN}}$ as corresponding to information entropy. Now, since information entropy also corresponds, \textit{ex hypothesi}, to \textit{S}$_{\text{TD}}$, we have an indirect argument for the correspondence of \textit{S}$_{\text{VN}}$ to \textit{S}$_{\text{TD}}$. 

However, \textbf{H$\&$S} claim that this argument fails for \textit{collapse} interpretations, i.e. interpretations of quantum mechanics on which a superposed quantum state ontologically collapses into a pure state upon measurement (either precisely or approximately).\footnote{On GRW-type approaches, though, collapse occurs with or without measurement, but measurement increases the likelihood of collapse, roughly speaking.} They allow that, on \textit{no-collapse} interpretations, e.g. Bohmian or many-worlds interpretations, the location measurement in stage five does not decrease \textit{S}$_{\text{VN}}$, since the state of the system never changes in light of measurements, and so the above argument goes through.

Let us see what they could mean by this claim by following the state of the particle through the cycle. At stage two, everyone agrees that the state of the system is $\rho_{spin}$ following the z-spin measurement; \textit{S}$_{\text{VN}}$ increases by $ln$ 2. At this point, the particle's position degrees of freedom remain independent from its spin degrees of freedom, as per our ideal gas assumption, though we might assume the particle starts out on the left half of the box, with the mixture of position states $\rho_{pos}(L)$ with `L' representing the left side. (Consider \textit{Fig. 1} but with only one particle). Following the semipermeable wall's filtering at stages three and four, the location of the particle becomes classically correlated with the spin. Let's say that the semipermeable wall sends $\ket{\psi_z^{\uparrow}}$ particles to the left, represented by $\rho_{pos}(L)$, and $\ket{\psi_z^{\downarrow}}$ particles to the right, represented by $\rho_{pos}(R)$. As such, the (mixed) state of the particle is now: 
\begin{equation}
    \rho_{particle} = \frac{1}{2}\bigg(\ket{\psi_z^\uparrow}\bra{\psi_z^\uparrow} \otimes \rho_{pos}(L) + \ket{\psi_z^\downarrow}\bra{\psi_z^\downarrow} \otimes \rho_{pos}(R)\bigg)
\end{equation}

\noindent For no-collapse interpretations, \textbf{H$\&$S} agree that the state of the particle stays the same as above after the location measurement in stage five. We perform the compression in stage five and remove the partition at the end of stage six, thereby removing the classical correlations between position and spin. No further change in either information entropy or \textit{S}$_{\text{VN}}$ occurs, and hence the correspondence goes through (\textbf{H$\&$S} 2006, 164) - the spin state remains mixed until unitarily transformed into a pure state and completing the cycle. 

For collapse interpretations, they claim that the location measurement \textit{decreases} \textit{S}$_{\text{VN}}$ by $ln$ 2, because, on collapse interpretations, the state of the particle upon the measurement, depending on which side the particle is found, becomes: 

\begin{equation}
        \rho_{particle}=
    \begin{cases}
      \ket{\psi_z^\uparrow}\bra{\psi_z^\uparrow} \otimes \rho_{pos}(L)  \\
      \ket{\psi_z^\downarrow}\bra{\psi_z^\downarrow} \otimes \rho_{pos}(R)
    \end{cases}
\end{equation}

\noindent The spin state of the system here effectively goes from being a mixed state to a pure state as a result of this measurement: \textit{S}$_{\text{VN}}$ \textit{decreases} by $ln$ 2. Summing up the entropy changes, there was a decrease of $ln$ 2 in information entropy, and a net change of zero for \textit{S}$_{\text{VN}}$ as a result of the increase in stage two and the decrease in stage five. Overall, then, the change is \textit{not} zero but $-ln$ $2$; our accounting has gone awry, and there is a failure of correspondence between \textit{S}$_{\text{VN}}$ and information entropy. If this is right, \textit{S}$_{\text{VN}}$ does not correspond to \textit{S}$_{\text{TD}}$.

However, I think that \textbf{H$\&$S} are wrong to claim that \textit{S}$_{\text{VN}}$ decreases following the location measurement for collapse interpretations. As Prunkl (Prunkl ms, 11--12) notes, there is an inconsistency here. Everyone, \textit{including} \textbf{H$\&$S}, agrees that the spin state of the particle is mixed -- \textit{not} pure -- after stage two's spin measurement, \textit{even on} collapse interpretations (\textbf{H$\&$S} 2006, 160). In that case, why does the particle's spin become \textit{pure} after the location measurement? 

I think this results from a confusion over the nature of mixed states. In particular, they seem to have adopted what Hughes (Hughes 1992, $\S$5.4, $\S$5.8) call the ``ignorance interpretation'' of mixed states, confusing what I call \textit{classical} and \textit{quantum} ignorance. They seem to be assuming that mixed states simply represents \textit{classical} ignorance, i.e. the lack of knowledge about a \textit{particular system}: a system represented by a mixed state \textit{really is} in a pure state, but we know not which. This is why the location measurement is supposed to reveal to us the pure state of this system (by revealing which side it is on and hence the correlated spin state) and hence `wash away' our classical ignorance of the real state of the system - post-measurement, we know exactly which pure state \textit{this} system is in, unlike pre-measurement; hence \textit{S}$_{\text{VN}}$ decreases. 

However, as Hughes (Hughes 1992, 144/150) argues, this interpretation of mixed states -- as representing classical ignorance about which pure state a \textit{particular} system is in -- cannot be the right interpretation of all mixed states. To begin, a mixed state can be decomposed in non-unique ways in general. Here's a simple example: a mixed state representing a mixture of $\ket{\psi_z^\uparrow}$ and $\ket{\psi_z^\downarrow}$ can \textit{also} represent a mixture of $\ket{\psi_x^\uparrow}$ and $\ket{\psi_x^\downarrow}$ and so on. If we insist that a mixed state represent our classical ignorance about the real state of a particular system, then we end up having to say that a system's state is really \textit{both} either $\ket{\psi_z^\uparrow}$ or $\ket{\psi_z^\downarrow}$, \textit{and} either $\ket{\psi_x^\uparrow}$ or $\ket{\psi_x^\downarrow}$. Of course, this is impossible given quantum mechanics. The defender of the classical ignorance interpretation might insist that we simply pick one pair of possible pure states but not both at once. In general, however, there's no way to do that non-arbitrarily given some density matrix. Furthermore, this problem only worsens when we consider that there are usually more than just two ways to decompose a density matrix - a principled choice based on the mixed state alone is not feasible. The mixed state cannot be a representation of classical ignorance.
 
Instead, to paraphrase Hughes (Hughes 1992, 144--145), mixed states should be (minimally) interpreted as such: if we prepared in \textit{the same way} an ensemble of systems, each described with the same mixed state, i.e. a mixture of pure states with certain weights, then the relative frequency of any given measurement outcome from the ensemble is exactly what we would get if the ensemble comprised of various `sub-ensembles' each in one of the pure states in the mixture, with the relative frequency of each sub-ensemble in the ensemble given by the corresponding weights. 

In other words, the sort of \textit{quantum} ignorance relevant in the right interpretation of mixed states is not whether we are ignorant about the \textit{real} state of \textit{this} particular system, but whether we are ignorant about the \textit{measured} states of an \textit{ensemble} of \textit{identically prepared} systems like this one. If this is right, quantum ignorance cannot be `washed away' upon measurement of a single system unlike the sort of ignorance \textbf{H$\&$S} were implicitly assuming, and it seems like this quantum ignorance is precisely what remains after the location measurement. 

This was roughly Henderson's (2003) criticism against Shenker (1999), which is why it is puzzling that \textbf{H$\&$S} (2006) commit the same mistake:

\begin{quote}
This preparation produces the pure states [$\ket{\psi_z^\downarrow}$] and [$\ket{\psi_z^\uparrow}$] with equal probabilities. In a particular trial, the observer may take note of the measurement result, and he therefore discovers that he has say a [$\ket{\psi_z^\uparrow}$]. If he applies a projective measurement in the [$\{\ket{\psi_z^\uparrow}, \ket{\psi_z^\downarrow}\}$] basis, he could predict that he will measure [$\ket{\psi_z^\uparrow}$]. However, this does not mean that, if someone handed him another state prepared in the same way, he could again predict that the outcome of his measurement would be [$\ket{\psi_z^\uparrow}$]. In this sense the observer does not know the state of the system which is being prepared, and it is because of this ignorance that the state is mixed. Looking at the measurement result does not remove the fact that there is a probability distribution over the possible outcomes. (Henderson 2003, 294)
\end{quote}

\noindent This applies to the location measurement in stage five too: measuring the location of the particle in \textit{this} case does not change the state of the particle from a mixed one to a pure one even on collapse interpretations. Firstly, it seems quite irrelevant whether we adopt a collapse or no-collapse interpretation, because the collapse mechanism applies to superposed pure states, not statistical mixtures. If anything, collapse had already happened in the stage two measurement procedure, yet everyone including \textbf{H$\&$S} (\textbf{H$\&$S} 2006, 160) accepts that the system is in a \textit{mixed} state after stage two even for collapse interpretations. More importantly, there remains a probability distribution over the states of the particle as a result of stage two's spin measurement, even \textit{after} the location measurement. Given an ensemble of particles prepared from stages one to five in the same way, we are \textit{still} not be able to predict with certainty whether an ensemble of particles would all be measured on the left or right sides of the box (and hence all spin-up or spin-down) as a result of the mixed state resulting from stage two, only that half of the ensembles will be on the left and the other half will be on the right. Quantum ignorance remains -- the system remains in a mixed state even after the location measurement, as: 

\begin{equation}
     \rho_{particle} = \frac{1}{2}\bigg(\ket{\psi_z^\uparrow}\bra{\psi_z^\uparrow} \otimes \rho_{pos}(L) + \ket{\psi_z^\downarrow}\bra{\psi_z^\downarrow} \otimes \rho_{pos}(R)\bigg)   
\end{equation}

\noindent This is exactly the state of the system in no-collapse interpretations, i.e. quantum ignorance does \textit{not} discern between collapse and no-collapse interpretations. What \textit{has} gone away is the classical ignorance that \textbf{H$\&$S} (mistakenly) assumed was relevant for mixed states, ignorance about\textit{ this particular system}'s state. By measuring the system's location, we come to learn of the correlations between location measurement and the particle's spin. This ignorance does not change the mixed state to a pure state: instead, this loss of classical ignorance -- \textit{gain in information} -- is represented as a \textit{decrease} in information entropy just as before, and this information is what we \textit{use} to perform the compression in stage five.

As a result, there is no additional decrease in \textit{S}$_{\text{VN}}$ in stage five for collapse interpretations; the entropy accounting lines up after all, as with no-collapse interpretations: the decrease in information entropy \textit{does} correspond to the increase in \textit{S}$_{\text{VN}}$, and so information entropy does correspond to \textit{S}$_{\text{VN}}$ after all. \textbf{H$\&$S}'s argument does not establish the failure of correspondence between \textit{S}$_{\text{VN}}$ and \textit{S}$_{\text{TD}}$ via the failure of \textit{S}$_{\text{VN}}$ and information entropy to correspond. 

To sum up, their arguments in the single-particle case are either ill-motivated and irrelevant to von Neumann and our discussion of correspondence when considered in terms of phenomenological thermodynamics, or outright fails when considered in the more relevant domain of (informational approaches to) statistical mechanics. Either way, their argument does not support the failure of correspondence between \textit{S}$_{\text{VN}}$ and \textit{S}$_{\text{TD}}$.\footnote{Let me briefly note that their argument in the two particles case fails for similar reasons. On the one hand, from the perspective of phenomenological thermodynamics, their argument is irrelevant: following Maxwell and others, two particles do not a thermodynamic system make. On the other hand, in the domain of statistical mechanics, the analysis in terms of information entropy is irrelevant from non-informational views of statistical mechanics. From an informational perspective, however, their argument rests again on the supposed difference between collapse and no-collapse interpretations of mixed states. Since this difference is non-existent, their argument likewise fails apart in that case.}

\subsection{Finitely Many Particles}

\textbf{H$\&$S}'s argument in the case of finitely many particles rests on the assumption of \textit{equidistribution}, i.e. that the particles will be equally distributed across the left and right sides of the box after separation by the semi-permeable wall. 

Assuming equidistribution, the increase in \textit{S}$_{\text{VN}}$ given the spin measurement in stage two is $N ln 2$ (\textbf{H$\&$S} 2006, 169). Furthermore, the decrease in thermodynamic entropy in the fourth stage is $N ln 2$ as well. The entropic accounting therefore seems to work out.

However, \textbf{H$\&$S} press further on the `rough' nature of equidistribution when $N$ is large but finite: they claim that the change in \textit{S}$_{\text{VN}}$ will only only be $N ln 2$ when $N$ is infinite, since equidistribution only truly holds when $N \rightarrow \infty$. In other cases, \textit{S}$_{\text{VN}}$ will strictly only \textit{approximate} \textit{S}$_{\text{TD}}$, and hence \textit{S}$_{\text{VN}}$ and \textit{S}$_{\text{TD}}$ combined will never be \textit{exactly} zero; hence, ``Von Neumann’s argument goes through as an approximation'' (\textbf{H$\&$S} 2006, 169). However, they claim that this state of affairs suggest, instead, that von Neumann's argument strictly \textit{fails}: ``[...] since Von Neumann’s argument is meant to establish a conceptual identity between the quantum mechanical entropy and thermodynamic entropy, we think that such an implication is mistaken [...] no matter how large N may be, as long as it is finite, the net change of entropy throughout the experiment will not be exactly zero." (\textbf{H$\&$S} 2006, 169)

\noindent As I have already discussed in $\S$4.1, it is not clear to me that von Neumann's goal really was to establish \textit{strict identity} (what they call ``conceptual identity''), i.e. correspondence between \textit{S}$_{\text{VN}}$ and \textit{S}$_{\text{TD}}$ in \textit{all} domains. Rather, it seems to be the establishing of correspondence \textit{only} in domains where \textit{S}$_{\text{TD}}$ is taken to hold. If so, their argument here simply misses the point.

Furthermore, as is well-known, the particle analogue of thermodynamics, statistical mechanics, become equivalent to phenomenological thermodynamics only when $N = \infty$, viz. when $N$ arrives at the thermodynamic limit. As such, to complain that \textit{S}$_{\text{VN}}$ does not match up to \textit{S}$_{\text{TD}}$ outside of this domain is to demand the unreasonable, since it is not clear that even statistical mechanics, the bona fide particle analogue of thermodynamics, can satisfy this demand. Since \textit{S}$_{\text{VN}}$ approximates \textit{S}$_{\text{TD}}$ the same way statistical mechanical entropies approximate \textit{S}$_{\text{TD}}$ (and becomes equivalent at $N = \infty$), and physicists generally accept that statistical mechanics corresponds to thermodynamics nevertheless, why should this problem of approximation be particularly problematic for \textit{S}$_{\text{VN}}$? I think \textbf{H$\&$S} take too seriously the notion of conceptual identity involved in von Neumann's thought-experiment to be \textit{strict} equality, though I suspect a better way to understand von Neumann's strategy is to understand \textit{S}$_{\text{VN}}$ as an approximation to \textit{S}$_{\text{TD}}$ that is more fundamental than \textit{S}$_{\text{TD}}$ in small $N$ cases, but becomes part of the \textit{S}$_{\text{TD}}$ calculus in domains where \textit{S}$_{\text{TD}}$ applies.

To have a case against \textit{S}$_{\text{VN}}$ as a quantum analogue of \textit{S}$_{\text{TD}}$ in the case of finitely many particles, \textbf{H$\&$S} must explain what exactly the problem is with approximations in \textit{this} case, if it has worked out so well for the case of statistical mechanics and thermodynamics. If not, they might just be ``taking thermodynamics too seriously'.\footnote{See Callender (2001).}

One might say something stronger: unless they can justify why we cannot use approximations at all in science, they do not have a case at all. As they note themselves, \textit{S}$_{\text{TD}}$ is itself only \textit{on average approximately} $-N ln 2$ (\textbf{H$\&$S} 2006, 169), only being equal to $-N ln 2$ when $N = \infty$. So, in fact, the approximate quantity of \textit{S}$_{\text{VN}}$, $\sim N ln 2$, exactly matches the approximate quantity of \textit{S}$_{\text{TD}}$, $\sim-N ln 2$, in the case of finitely many particles. Unless there is something wrong with approximations in physics \textit{in general}, this, then, is in fact a case of \textit{S}$_{\text{VN}}$ corresponding to \textit{S}$_{\text{TD}}$, contrary to their argument.

\subsection{Infinitely Many Particles}

\textbf{H$\&$S} consider von Neumann's argument in the infinite particles case in two different ways: one as $N \rightarrow \infty$ and one as $N = \infty$. As they rightly point out, the two cases are very different for calculations of physical quantities.

Consider stage two and stage five in this context. \textbf{H$\&$S} emphasize that a spin measurement is ``a physical operation which takes place in time'' (\textbf{H$\&$S} 2006, 170), which constrains what is physically possible. 

For the case where $N \rightarrow \infty$, stage two is to be understood as a succession of physical measurements where ``we measure individual quantities of each of the particles separately and only then count the relative frequencies'' (\textbf{H$\&$S} 2006, 170), before coming up with a density matrix describing this state. In this case, as with the case described in $\S$4.3, \textit{S}$_{\text{VN}}$ approaches $N ln 2$ as $N \rightarrow \infty$. Their complaint here consist of two premises: one, that, as with $\S$4.3, \textit{S}$_{\text{VN}}$ never reaches $N ln 2$ unless $N = \infty$. Two, that since measurements are physical measurements, we can never perform an infinite series of these measurements, and so we can never measure infinite particles. \textit{A fortiori} the measurable \textit{S}$_{\text{VN}}$ can never arrive at $N ln 2$, and so the entropic accounting is again supposed to be inconsistent if we consider both \textit{S}$_{\text{VN}}$ and \textit{S}$_{\text{TD}}$. 

However, it is clear that their argument is moot given a clear understanding of the sort of thermodynamics we are interested in (see $\S$4.3). While it is true that \textit{S}$_{\text{VN}}$ will never reach $N ln 2$, recall that \textit{S}$_{\text{TD}}$ (or, more likely, one of its statistical mechanical analogues, given the domain of finitely many particles merely \textit{approaching} $\infty$ rather than $N = \infty$) will likewise never reach $N ln 2$. In other words, it does not matter that we can never perform an infinite series of these measurements, and hence never come to know of \textit{S}$_{\text{VN}}$ at the thermodynamic limit, since we can likewise never have a thermodynamic entropy equivalent to $N ln 2$ unless we are at the thermodynamic limit. The two entropies, then, in fact \textit{correspond} in this case.

What of the second case? Here, \textbf{H$\&$S} concede that ``arithmetically Von Neumann’s argument goes through at the infinite limit'' (\textbf{H$\&$S} 2006, 172), which makes sense because, as I have insisted so far, von Neumann's strategy was never to demonstrate the \textit{strict identity} of \textit{S}$_{\text{VN}}$ and \textit{S}$_{\text{TD}}$, i.e. the correspondence of \textit{S}$_{\text{VN}}$ and \textit{S}$_{\text{TD}}$ in \textit{all} domains. Instead, it was to show that \textit{S}$_{\text{VN}}$ corresponds to \textit{S}$_{\text{TD}}$ \textit{only in the domain where phenomenological thermodynamics hold}, in all other cases merely \textit{approximating} \textit{S}$_{\text{TD}}$ in large $N$ cases or replacing it altogether (in e.g. single-particle cases). I maintain that \textbf{H$\&$S}'s main mistake was to confuse the domain where phenomenological thermodynamics hold, with domains where they do not hold.

\textbf{H$\&$S} complain that ``[...] real physical systems are finite. This means that Von Neumann’s argument does not establish a conceptual identity between the Von Neumann entropy and thermodynamic entropy of physical systems. Identities of physical properties mean that the two quantities refer to the same magnitude in the world." (\textbf{H$\&$S} 2006, 172) In line with what I have said in $\S$4.1, it seems that there was no physically meaningful theoretical term in phenomenological thermodynamics that \textit{could} refer to some quantity in the single-particle case, which was why von Neumann needed to come up with a new measure of entropy to begin with. Furthermore, extending a concept to a new domain does not require strict identity, as we have seen and understood for a long time in the case of statistical mechanics and phenomenological thermodynamics. 

As Peres (2002) summarizes: ``There should be no doubt that von Neumann’s entropy. . . is equivalent to the entropy of classical thermodynamics. (This statement must be understood with the same vague meaning as when we say that the quantum notions of energy, momentum, angular momentum, etc., are equivalent to the classical notions bearing the same names)." (Peres 2002, 174) `Equivalence' here should not be understood in terms of strict (or conceptual) identity i.e. correspondence at all domains. Rather, we should understand equivalence loosely as correspondence in the suitable domains of application, and successful extension of old concepts in these domains to new domains. As Peres noted above, `equivalence' should be understood in the context of discovery, where one is trying to develop new concepts which are analogous to old ones in different domains. For von Neumann, we have a theory (phenomenological thermodynamics) that is well-understood, but also another theory (quantum mechanics) that we want to understand in light of the former theory. Finding correspondence provides us with ways to \textit{extend} concepts from the original theory to the new theory: for example, with \textit{S}$_{\text{VN}}$ we may now define `something like' \textit{S}$_{\text{TD}}$ whereas before there was no way to talk about these cases. The same goes for statistical mechanics: by finding a correspondence between e.g. temperature to mean kinetic energy in the thermodynamic limit, we can \textit{extend} the notion of `something like' temperature beyond its original domain into systems with small numbers of particles, whereas before there was, again, no way to talk about these cases.  

I see nothing wrong in these cases in the context of discovery. We should give up a strong and untenable notion of conceptual identity in this context. If so, \textbf{H$\&$S}'s objection loses much bite. 

They further claim that ``the fact that the behavior of the two quantities coincides approximately for a very large number of particles is not enough, because in any ensemble of finite gases there are systems in which the identity will not be true. This means that in a real experiment the Von Neumann entropy is not identical with the thermodynamic entropy." (\textbf{H$\&$S} 2006, 172) This again reveals a confusion between phenomenological and statistical thermodynamics. If they want to talk about particles \textit{at all}, it seems they must adopt some form of statistical mechanical picture with microscopic variables, given phenomenological thermodynamics' emphasis on purely macroscopic variables like volume or temperature. Yet, if so, they must recognize that thermodynamic entropy \textit{S}$_{\text{TD}}$ is in general not strictly identical to statistical mechanical entropy, e.g. the Gibbs entropy or information entropy (briefly discussed in $\S$4.2) \textit{either}. Their complaint about approximate coincidences not being enough for (the relevant sort of) equivalence thus weakens significantly, especially since they must assume some such equivalence (which cannot be strict identity) to even talk about particles within the context of phenomenological thermodynamics to begin with. Furthermore, statistical mechanics is evidently empirically successful in explaining and predicting traditionally thermodynamic phenomena despite this `non-equivalence' -- it is not clear why this `non-equivalence' should matter if, for all practical purposes, statistical mechanics is the conceptual successor of thermodynamics. Of course, if they could come up with a principled reason why approximations should not be allowed \textit{period}, \textit{while} accounting for statistical mechanics' empirical success in accounting for thermodynamic behavior, then this could change. As of now, I see no such argument forthcoming.

\section{Conclusion, and Some Open Questions}

Given the above, I hope to have shown that \textbf{H$\&$S}'s argument against the correspondence of \textit{S}$_{\text{VN}}$ and \textit{S}$_{\text{TD}}$  -- to my knowledge the only one in the philosophical literature -- fails to hold in all three cases considered ($\S$4.1 -- $\S$4.4), as a result of their misunderstanding about domains where phenomenological thermodynamics should hold and domains where it should not. This is compounded with misunderstandings about the role of approximations and the relevant interpretation of density matrices and ignorance in quantum mechanics. I conclude that their argument fails on the whole; the correspondence holds for now. 

Of course, even if \textbf{H$\&$S}'s claims were debunked, this does not yet amount to a positive argument for the equivalence between von Neumann entropy and thermodynamic entropy. Even assuming correspondence, correspondence does not entail equivalence. However, the former does provide good \textit{prima facie} reasons to believe the latter, especially given the novel take on correspondence I provided in the end of $\S$3: we can accept the correspondence based on thermodynamic considerations about the Second Law and \textit{S}$_{\text{TD}}$ accounting, but \textit{also} based on quantum mechanical considerations about \textit{S}$_{\text{VN}}$ accounting. The correspondence supports a `two-way street' -- equivalence -- between \textit{S}$_{\text{TD}}$ and \textit{S}$_{\text{VN}}$. 

While I hope to have conclusively refuted \textbf{H$\&$S}'s argument, this is but the beginning of further inquiry into questions arising from this supposed correspondence. Amidst the tangle of entropies, there remains much more housekeeping to be done for philosophers of physics.

\newpage
\nocite{*}
\bibliography{bib.bib}

\end{document}